# Homogeneous Charge Relaxation in Organic Mixed Ionic-Electronic Conductors

**Authors:** Silan Zhang[1, 2, 3], Scott T. Keene[2, 3], Junfu Tian[5], Georgina E. Lindop[4], Simon M. Fairclough[4], Zhengkang Qu[3], Xian Wei Chua[3,6], Christopher Slaughter[1, 2], Christina J. Kousseff[5], Caterina Ducati[4], Iain McCulloch[5,7], Henning Sirringhaus[3], Akshay Rao[3*], George G. Malliaras[1, 2*]

**Affiliations:**

[1] Institute for Biomedical Innovation, Department of Engineering, University of Cambridge, Cambridge, UK.

[2] Electrical Engineering Division, Department of Engineering, University of Cambridge, Cambridge, UK.

[3] Cavendish Laboratory, University of Cambridge, Cambridge, UK.

[4] Department of Materials Science and Metallurgy, University of Cambridge, Cambridge, UK

[5] Department of Chemistry, University of Oxford, Oxford, UK.

[6] Department of Chemical Engineering and Biotechnology, University of Cambridge, Cambridge, UK.

[7] Andlinger Center for Energy and the Environment and Department of Electrical and Computer Engineering, Princeton University, Princeton, NJ, 08544, USA.

*Corresponding author: Email: ar525@cam.ac.uk (Akshay Rao); gm603@cam.ac.uk (George G. Malliaras)

**Abstract:** Organic mixed ionic-electronic conductors (OMIECs) couple ionic and electronic signals, enabling energy storage, neuromorphic computing, and bioelectronics. However, how charge relaxes in water once the bias is removed, which determines how long stored states persist, remains unclear. Using operando charge photometry, we tracked charge in OMIEC films during biasing and open-circuit relaxation. Under bias, charges moved as redox fronts. After bias removal, charge instead relaxed uniformly across the channel. Modeling and cryo-electron microscopy indicate that swelling creates connected water-rich pathways through the bulk. These pathways speed ion motion but also let stored charge leak away. Tuning the terminal side chain of a polythiophene controls water uptake and ion trapping. Bulky hydrophobic side chains enable stable programmable states, whereas hydrophilic side chains support fast electrophysiology sensing.

## Main Text:

Organic mixed ionic-electronic conductors (OMIECs) have emerged as a foundational material class for bridging ionic and electronic signals, which underlie the function of a range of technologies, from high-rate energy storage devices to low-power neuromorphic computing(*1-3*). By allowing ions from a biological medium to penetrate their bulk, OMIECs act as a seamless bridge between biological and electronic systems. This volumetric charge coupling(*4-8*) allows ionic flux to modulate the internal electronic charge carrier density, effectively transducing biological signals into high-gain electronic responses(*9*). However, despite their increasingly widespread application, the fundamental dynamics of these coupled charges, particularly in hydrated environments, are not well understood(*10*).

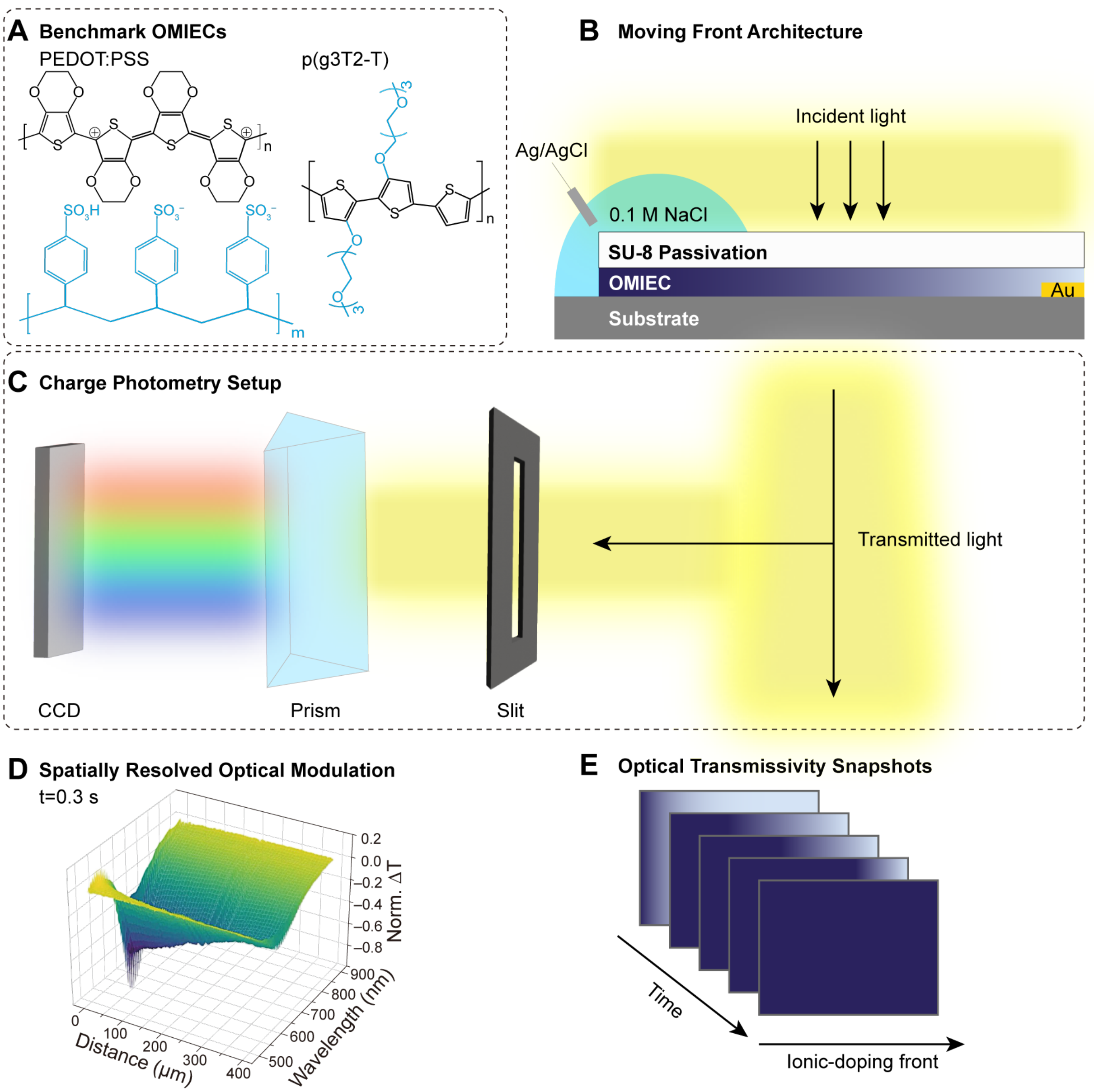

**Fig. 1. Operando charge photometry (CP) for resolving OMIEC transport dynamics. (A)** Chemical structures of the benchmark OMIECs PEDOT:PSS and p(g3T2-T). **(B)** Moving front

device architecture. The OMIEC layer is sandwiched between a transparent substrate and an SU-8 passivation layer. By contacting the film with a spatially separated electrode (electronic source) and a 0.1 M NaCl electrolyte (ionic source), this geometry enables the spatial deconvolution of ionic and electronic transport as they propagate along the polymer channel. **(C)** Charge photometry. The optical system is configured for dual-mode operando detection: transmitted light is either directly captured to produce **(E)** or directed through a precision slit and dispersed by a prism to generate **(D)**. **(D)** Spatially resolved optical spectroscopy, presented as normalized differential transmittance (Norm. ΔT) as a function of distance. Representative data shows the electrochemical de-doping of PEDOT:PSS (0 to 0.8 V vs. Ag/AgCl) at t = 0.3 s. **(E)** Time-resolved Norm. ΔT transmittance snapshots captured during operando characterization to visualize real-time charge distribution across the channel.

Current design strategies for OMIECs typically integrate conjugated polymer backbones for electronic conduction with hydrophilic sidechains to facilitate ion transport(*11*). For instance, as shown in **Fig. 1A**, the commercially available poly(3,4-ethylenedioxythiophene):poly(styrene sulfonate) (PEDOT:PSS) utilizes a water-soluble polyanion ($PSS^{-}$) blend, while widely used glycolated polythiophenes such as p(g3T2-T) (see chemical structure in **Fig. 1A**) incorporate ethylene glycol sidechains to promote ion permeability(*12*). In these disordered systems, electronic charge transport proceeds through thermally activated hopping between states, while ionic species migrate via hopping coupled with segmental chain motion in dry states, transitioning to rapid transport as solvated ions within the swollen, hydrated matrix(*13*). Upon immersion in aqueous media, these hydrophilic components allow material swelling, creating a hydrated matrix that facilitates volumetric ion penetration. Modulating these swelling properties via side chain engineering is known to influence response speed and operation stability(*14*). However, how hydration governs ion transport and charge redistribution within OMIECs remains poorly understood. Answering this question is crucial, as the optimized operation of these materials within swollen hydrated environments is key to their future use.

A particularly overlooked regime is open-circuit relaxation, that is, how the injected charges evolve without external driving. From a fundamental perspective, this regime isolates the intrinsic transport behavior of the material: under bias, the applied field and electronic contact dictate where and how fast charges move, whereas at open circuit the evolution is governed by the internal ionic pathways, ionic-electronic coupling and hydration state of the film. From an application perspective, open-circuit relaxation determines how long a programmed state persists. It sets the retention of conductance states in neuromorphic devices after the programming bias is removed(*3*), the self-discharge of OMIEC-based charge-storage devices(*15*), and the recovery and drift of electrochemical sensors and bioelectronic interfaces between successive operations(*16*). Despite its importance, relaxation is typically characterized only through device-level metrics such as open-circuit potential or electrical signal decay, which cannot reveal where within the film charge relaxes.

Characterizing mixed electronic-ionic transport within the swollen hydrated environment is extremely challenging. Techniques such as quartz crystal microbalance (QCM-D) and atomic force microscopy (AFM) have quantified macroscopic swelling and mechanical shifts, yet they remain limited in resolving underlying microstructural dynamics(*17-19*). Operando grazing-incidence wide-angle X-ray scattering (GIWAXS) has provided critical insights into crystalline transitions and lattice expansion during electrochemical doping(*20, 21*). Complementarily, cryogenic scanning transmission electron microscopy (cryo-STEM) has enabled the visualization of nanoscale morphology and disorder-to-order transformations that occur within the polymer matrix(*22*). While operando nuclear magnetic resonance (NMR) spectroscopy has further succeeded in tracking cation-coupled water migration(*23*), directly visualizing where and when charge-state changes occur across an OMIEC film during operation remains challenging. This is a missing link in understanding how OMIECs operate.

To address this gap, we use operando charge photometry (CP) to track where and when charges move inside hydrated OMIEC films. CP tracks local optical signatures of the polymer during device operation, allowing us to compare charge transport under an applied bias with charge relaxation after the bias is removed, the regime that governs state retention and self-discharge. Using this approach, we first confirm that under bias, charge moves as a front propagating, consistent with classical drift-diffusion. However, once the bias is removed, both PEDOT:PSS and p(g3T2-T) relax almost uniformly across the entire film rather than from the interface. We label this behavior homogeneous charge relaxation. A transport model based on Nernst–Planck ion motion reproduces this behavior when hydration-enhanced ionic transport is included, and cryogenic electron microscopy reveals water-rich regions and hydration-induced structural expansion in PEDOT:PSS. These findings suggest that water plays a dual role in OMIECs. It enables fast ion motion during operation, but also provides pathways for ions to leak back into the electrolyte at open circuit. We then use this insight to design two related polythiophene OMIECs with the same conjugated backbone but different terminal side chains. A bulky hydrophobic side chain limits water uptake and retains injected ions to hold programmable conductance states for non-volatile memory operation, whereas a hydrophilic side chain allows fast, reversible ion transport for physiological signal sensing.

## Operando charge photometry of OMIEC dynamics

To elucidate the complex charge dynamics in OMIECs, we utilized a moving front architecture(*24*) that allows for the spatial deconvolution of ionic and electronic transport (**Fig. 1B**). The device consists of a 400-µm-long OMIEC channel which is sandwiched between a transparent glass substrate and SU-8 ion-blocking passivation layer. One end of the channel is in contact with an aqueous electrolyte (0.1 M NaCl), serving as the ionic source, while the opposite end is connected to a gold (Au) electrode as the electronic contact. Upon applying an external bias between the Au and an Ag/AgCl reference electrode, ions are driven into the polymer bulk to maintain charge neutrality as the electronic carrier density is modulated. This electrochemical (de)doping process significantly alters the polymer's optical absorption profiles, leading to changes in its spectroscopic transmittance.

We monitored these operando optical transitions using a custom-built charge photometry setup (**Fig. 1C**). The incident light penetrates the active area of the moving device and is directed through two distinct detection pathways. In the spectroscopic mode, the transmitted light is passed through a precision slit and dispersed by a prism, yielding spatially resolved, normalized transmittance changes (Norm. ΔT) across the film. **Fig. 1D** presents a representative 3D spectroscopic map at t = 0.3 s during de-doping, capturing the local electronic state at each spatial coordinate. Simultaneously, the imaging mode captures time-resolved snapshots of the entire channel **(Fig.**

**1E**). In the PEDOT:PSS system, cation injection drives a visible dedoping boundary that propagates from the electrolyte interface toward the electronic contact—a behavior consistent with drift-diffusion models.

## Homogeneous charge relaxation in OMIECs

To monitor the electronic state changes during device operation, we obtained spectra at a fixed spatial position (x=50 µm from the electrolyte interface). Upon applying a potential of –0.8 V vs Ag/AgCl, the injection of $Na^+$ into PEDOT:PSS drives the electrochemical reduction of the polymer backbone. As shown in **Fig. 2A**, this process is manifested as a progressive decrease in the normalized transmittance ($\Delta T/T_{\max}$), with a spectral peak centered at ~682 nm. This feature corresponds to the recovery of the neutral π-π* transition, a signature of the PEDOT chains transitioning from a conductive doped state to a neutral insulating state(*25*). In parallel, we tracked the spatial transmittance changes at a fixed wavelength of 682 nm (**Fig. 2B**). The dedoping of PEDOT:PSS initiated at the electrolyte interface and then spreads along the PEDOT:PSS film towards the Au electrode side (see **Supplementary video 1**), resulting in a moving front boundary.

Returning the potential to 0 V vs. Ag/AgCl drove the electrochemical redoping of PEDOT:PSS. At the fixed measurement position, the π-π* adsorption progressively decreased as the polymer returned to its oxidized state (**Fig. 2C**). The corresponding spatial profiles in **Fig. 2D** and **Supplementary video 2** reveal a strong position-dependent response. The optical signal first recovered at the electrolyte-facing edge, while regions farther along the channel remained in the dedoped state. A distinct transition region therefore separated the redoped and dedoped portions of the film. With time, this transition region shifted continuously toward the Au contact, producing a pronounced spatial gradient and a clear propagating redox front. Thus, during both bias-driven dedoping and redoping, the polymer redox state changes sequentially along the channel rather than simultaneously throughout the film.

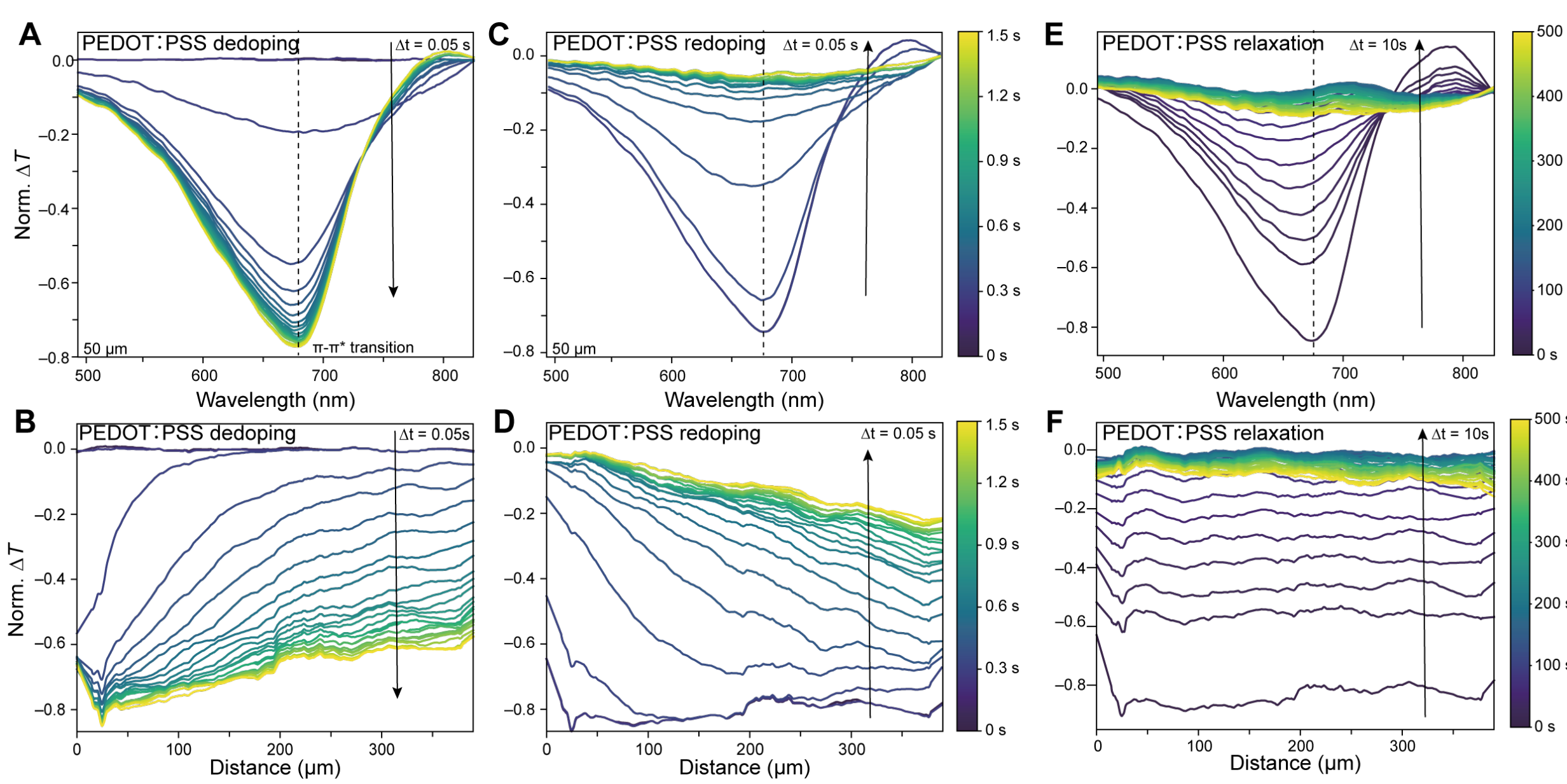

**Fig. 2. Operando visualization of propagating redox fronts and homogeneous charge relaxation in PEDOT:PSS. (A, C, and E)** Normalized differential-transmittance (Norm. ΔT)

spectra acquired at a fixed position 50 μm from the OMIEC–electrolyte interface during electrochemical dedoping **(A)**, redoping **(C)**, and subsequent open-circuit relaxation **(E)**. Spectra were collected at 0.05-s intervals during dedoping and redoping and at 10-s intervals during relaxation. The evolution of the π–π* transition (dashed line) tracks changes in the polymer redox state. **(B** and **D)** Spatiotemporal maps acquired during dedoping **(B)** and redoping **(D)** at 0.05-s intervals. Both processes exhibit propagating redox fronts and spatial concentration gradients, consistent with drift–diffusion transport. **(F)** Spatiotemporal maps acquired at 10-s intervals during open-circuit relaxation, showing spatially homogeneous recovery of the polymer redox state.

Integration of the charging current indicated that the fully dedoped PEDOT:PSS film contained approximately 0.3 mol $L^{-1}$ $Na^+$, about three times the concentration in the external electrolyte (**Fig. S1**). This concentration difference led us to expect that, after electrically disconnecting the Au and Ag/AgCl electrodes, $Na^+$ would leave first near the electrolyte interface and generate a relaxation front.

We then removed the external bias and monitored the film under the open-circuit conditions. At the fixed measurement position, the π-π* feature gradually decayed with time, indicating spontaneous reoxidation of PEDOT:PSS, or self-redoping (**Fig 2E** and **Fig. S2A** with 0.1 s interval). Because the $Na^+$ concentration in the dedoped film exceeded that in the external electrolyte, we initially expected ion release and polymer reoxidation to begin near the electrolyte-facing edge and then propagate along the channel.

The spatial evolution in **Fig. 2F** or **Fig. S2B** (with 0.1 s interval) and **Supplementary video 3** was qualitatively different from the redoping driven at 0 V vs. Ag/AgCl shown in **Fig. 2D**. No transition region emerged from the electrolyte interface, and no boundary translated toward the Au contact. Instead, the optical signal began to recover over nearly the entire channel. Successive spatial profiles changed approximately in parallel: their overall amplitude decreased with time, whereas the positions of the residual spatial features remained largely unchanged. Thus, the dominant evolution occurred in signal amplitude throughout the channel rather than through the displacement of a redox boundary. In addition to this distinct spatial behavior, the open-circuit relaxation occurred on a much longer timescale than the electrochemically driven redoping. This slow, channel-wide recovery was accompanied by a concurrent drift of the open-circuit potential (**Fig. S3**), indicating that the spatially uniform relaxation directly underlies device self-discharge.

Because the charge photometry signal tracks the local oxidation state of PEDOT:PSS, this non-propagating optical response indicates that the electronic charge density recovers nearly uniformly across the channel. Under local electroneutrality, the compensating ionic charge must therefore redistribute concurrently. The relaxation therefore reflects coupled ionic and electronic re-equilibration: as $Na^+$ redistributes toward the electrolyte, the PEDOT backbone is reoxidized and the electronic population re-equilibrates within the disordered density of states(*26*), producing the gradual decay of the π–π* optical signature.

Previous studies have shown that small anionic counterions can remain trapped in processed PEDOT:PSS films(*27*). To verify whether such trapped anions contribute to the homogeneous charge relaxation, we replaced the NaCl electrolyte with a PSSNa hydrogel, in which the polyanionic $PSS^-$ is immobilized and cannot penetrate the film. Homogeneous relaxation was still observed (**Fig. S4**), indicating that the effect does not require counterions. We further test the effect of film thickness by increasing the PEDOT:PSS thickness threefold to 540 nm (**Fig. S5**). The thicker film relaxed more slowly, retaining 20% of the initial optical response after 500 s, whereas the thinner film had nearly fully recovered. Despite this slower kinetics, the relaxation remained spatially homogeneous. Thus, film thickness influences the relaxation timescale but not the characteristic non-propagating spatial mode.

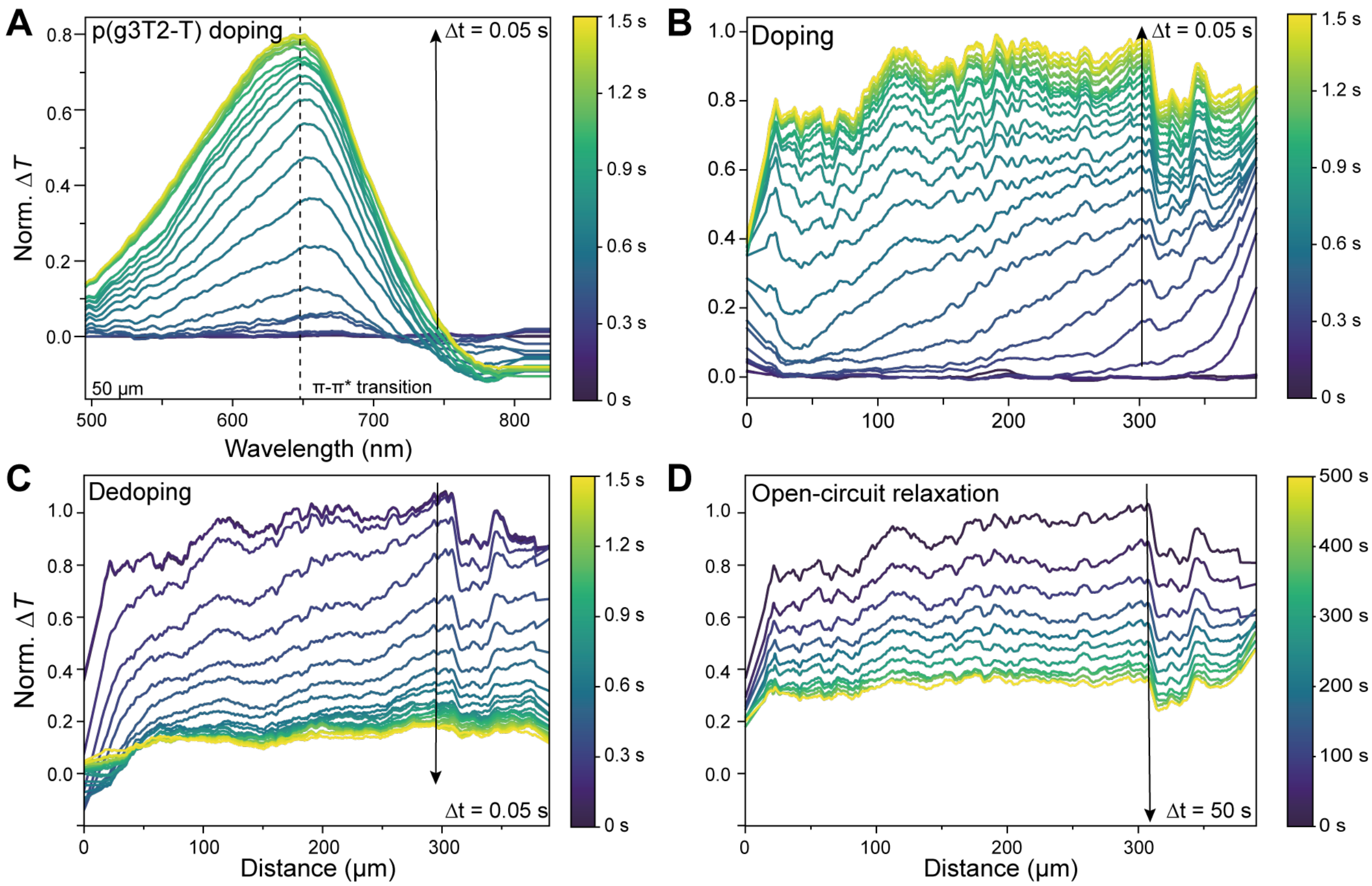

**Fig. 3. Propagating redox fronts and homogeneous charge relaxation in p(g3T2-T). (A)** Normalized differential-transmittance (Norm. ΔT) spectra measured 50 µm from the p(g3T2-T)–electrolyte interface during electrochemical doping at +0.6 V vs, Ag/AgCl . The bleaching of the π-π* transition near 650 nm is used to track the local oxidation state. **(B)** Spatiotemporal evolution of Norm. ΔT at 650 nm during voltage-driven doping, showing a propagating redox front **(C)** Spatiotemporal evolution of Norm. ΔT at 650 nm during voltage-driven dedoping at –0.4 V vs. Ag/AgCl, again showing a propagating redox front. **(D)** Spatiotemporal evolution of Norm. ΔT at 650 nm during subsequent open-circuit relaxation, showing a non-propagating, spatially homogeneous recovery across the channel.

We then questioned whether this phenomenon is unique to PEDOT:PSS due to its high pristine conductivity. To investigate this, we used a glycolated polythiophene, p(g3T2-T), which is non-conductive in its pristine state. When applying 0.6 V vs. Ag/AgCl, p(g3T2-T) was doped by $Cl^-$ from the electrolyte. The π-π* transition (at ~650 nm) was bleached, showing a positive transmission change(*28*) – the opposite as the PEDOT:PSS (**Fig. 3A**).

Spatially, however, the doping did not begin at the electrolyte-facing edge. Instead, the Norm. ΔT signal first increased near the Au contact and a distinct transition region formed from the Au contact toward the electrolyte interface, producing a propagating redox front (**Fig. 3B**). This propagation direction is consistent with hole-limited electrochemical doping at low carrier density, as reported previously(*9*).

We next reversed the bias to –0.4 V vs. Ag/AgCl to drive electrochemical dedoping. The corresponding spectra returned toward the undoped baseline (**Fig. S6**). In contrast to the doping process, the spatial response now initiated at the electrolyte interface: the Norm. ΔT signal decreased first near the electrolyte interface edge, while regions farther along the channel remained doped (**Fig. 3C**). The transition region propagated toward the Au contact. Thus, voltage-driven ion insertion and extraction both produced moving redox fronts, but with opposite propagation directions.

The behavior was completely different when the doped polymer was instead allowed to relax at open circuit. Spectroscopically, the π-π* bleaching gradually recovered with time, consistent with spontaneous self-dedoping (**Fig. S7A**). Spatially, however, no transition region emerged from either interface and no boundary propagated across the channel (**Fig. 3D** and **Fig. S7B**). Instead, the optical signal remained uniform across the channel while their overall amplitude decreased with time. The open-circuit response therefore reflects a non-propagating, spatially homogeneous relaxation of the polymer redox state.

Together with the PEDOT:PSS results, these measurements show that homogeneous charge relaxation is not specific to the highly conductive PEDOT:PSS system, but is also observed in a glycolated polythiophene with a distinct initial electronic state and exhibiting hole limited electrochemical doping.

### Origin of the homogeneous relaxation

We next asked what physical process could allow charge relaxation to occur across the entire channel without the propagation of a redox front. Because charge photometry reports the local polymer oxidation state, the spatially homogeneous optical response shows that the electronic charge density changes throughout the film during open-circuit relaxation. Under local electroneutrality, the compensating ionic charge must evolve concurrently.

We modeled $Na^+$ transport using a one-dimensional Nernst–Planck framework (**Fig. 4A, B,** See **Supplementary Text 1** for details)(*29*) in which the effective ion diffusivity depends on the electrolyte-accessible hydrated fraction of the PEDOT:PSS film . In the absence of electrolyte-accessible water pathways, the simulation produced a pronounced spatial concentration gradient: $Na^+$ was rapidly depleted near the electrolyte/polymer interface, whereas ions farther inside the channel remained at higher concentrations (**Fig. 4A**). The resulting profiles therefore evolved from the electrolyte interface edge and retained a strong spatial dependence, inconsistent with the nearly parallel relaxation observed experimentally.

We then introduced electrolyte-accessible hydrated regions into the model. Under this condition, the $Na^+$ concentration decreased throughout the channel with a much weaker spatial gradient (**Fig. 4B**). Rather than being decreased from the exposed electrolyte interface, ions could exchange through the hydrated fraction of the film, producing more spatially homogeneous relaxation. The model therefore shows that electrolyte-accessible water pathways are sufficient to reproduce the principal spatial characteristic of the experimental open-circuit response.

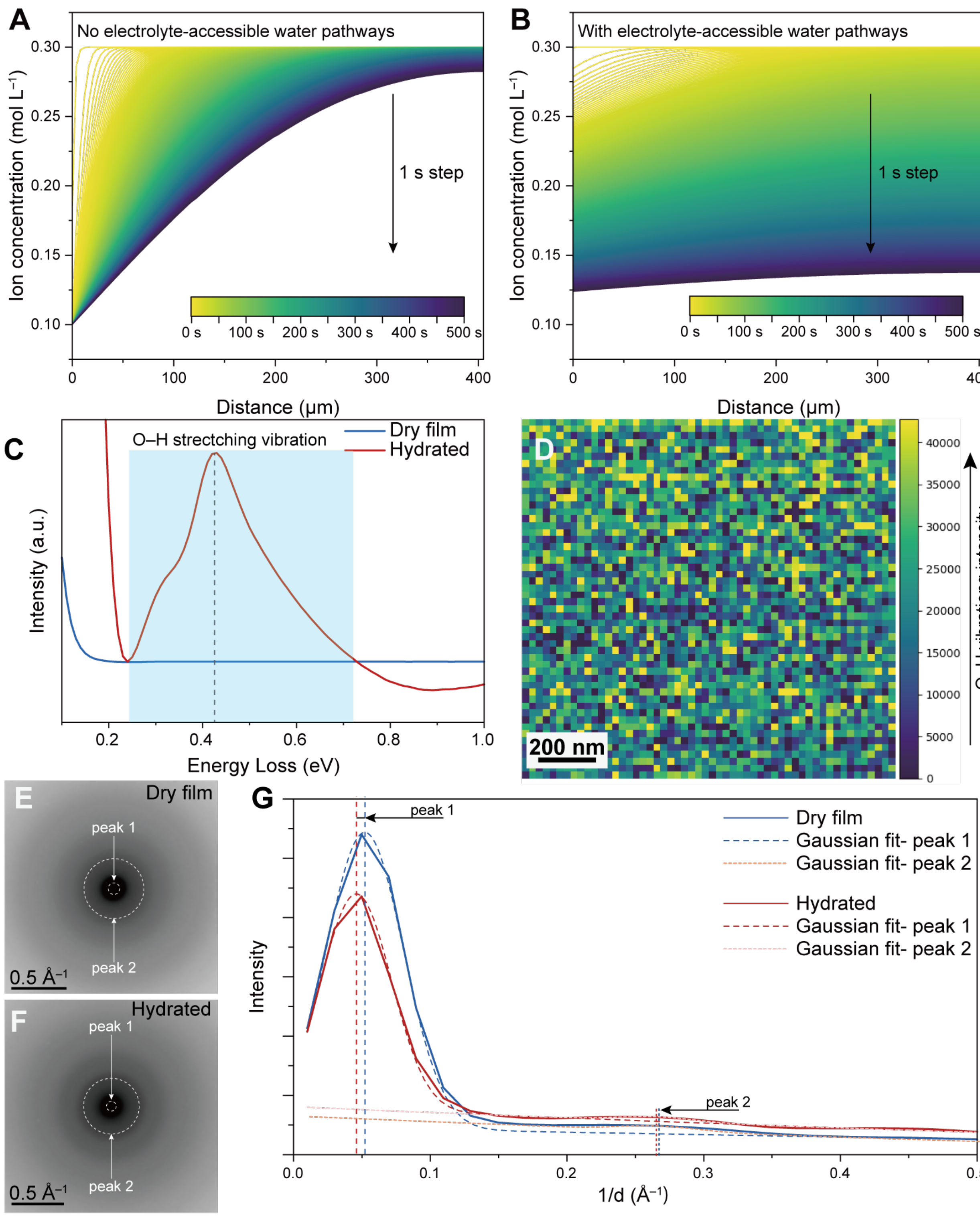

**Fig. 4. Electrolyte-accessible water pathways promote homogeneous charge redistribution in hydrated PEDOT:PSS. (A and B)** Simulated $Na^+$ concentration profiles during open-circuit relaxation using a one-dimensional Nernst–Planck transport model. **(A)** In the absence of electrolyte-accessible water pathways where the effective diffusion coefficient ($D_{\mathrm{effective}}$) for Na+ is $7 \times 10^{-11}$ m$^2$ s$^{-1}$, relaxation produces a pronounced spatial concentration gradient extending from the electrolyte/PEDOT:PSS interface. **(B)** Including electrolyte-accessible hydrated regions

corresponding to 35% swelling with $D_{effective}$ = 5.1 × $10^{-10}$ $m^2$ $s^{-1}$ produces substantially more spatially uniform $Na^+$ redistribution across the channel. Distance = 0 corresponds to the electrolyte/PEDOT:PSS interface. The corresponding effective $Na^+$ diffusion coefficients are derivate from the electrolyte-accessible volume fraction described in **Supplementary Text 1.** **(C)** Cryogenic vibrational electron energy-loss spectra of dry and hydrated PEDOT:PSS. Hydrated PEDOT:PSS exhibits an O–H stretching feature near 428 meV that is absent in the dry film. The shaded region indicates the energy-loss window integrated to generate the map in (D). **(D)** Two-dimensional map of the integrated O–H vibrational intensity over the energy-loss window indicated in (C), revealing a spatially heterogeneous distribution of water within the film.

**(E and F)** Summed scanning electron diffraction patterns over 1 μ$m^2$ area of dry **(E)** and hydrated **(F)** PEDOT:PSS.

**(G)** Azimuthally integrated diffraction profiles and Gaussian fits. Hydration shifts the first diffraction feature from a spacing of ~19.15 Å to ~21.72 Å and the PEDOT π–π stacking feature from ~3.68 Å to ~3.77 Å, indicating expansion of the PEDOT–PSS packing while largely preserving PEDOT π-stacking.

We further support the presence of water-rich pathways using cryogenic scanning transmission electron microscopy (cryo-STEM) coupled with electron energy loss spectroscopy (EELS). As shown in **Fig. 4C,** the hydrated film exhibits a pronounced O–H stretching vibration centered near 428 meV, arising from water retained within the film (*30*).  This feature is absent in dry PEDOT:PSS films. In addition to the main O–H vibration, the hydrated spectrum contains a shoulder near 320 meV (~2580 $cm^{-1}$), which may be associated with protonated O–H stretching in the acidic environment created by PSS (*31, 32*).

To determine how this water is distributed spatially, we integrated the vibrational EELS signal over the O–H stretching region highlighted in **Fig. 4C** and mapped the resulting intensity across the hydrated film (**Fig. 4D**). The O–H signal is strongly heterogeneous on the nanoscale: regions of relatively weak intensity are interspersed with, and surrounded by, regions with substantially stronger O–H signal. This distribution is consistent with the granular morphology previously reported for PEDOT:PSS, in which relatively hydrophobic PEDOT-rich regions are embedded within a more hydrophilic, water-affine PSS-rich matrix (*33-35*). Conversely, the dry PEDOT:PSS shows neither the pronounced O–H stretching feature from the EELS (**Fig. 4C**) nor O–H vibrational intensity in the corresponding two-dimensional map (**Fig. S8**).

We further examined how hydration affects the molecular packing of PEDOT:PSS using scanning electron diffraction (SED). The summed two-dimensional diffraction patterns of dry and hydrated PEDOT:PSS films are shown in **Fig. 4E** and **4F**, respectively. Both films exhibit similar diffraction features, indicating that the microstructure of PEDOT:PSS is retained after hydration. To compare the structural changes quantitatively, we azimuthally integrated the diffraction patterns to obtain one-dimensional intensity profiles (**Fig. 4G**). The first diffraction feature, corresponding to a spacing of ~19.15 Å in the dry film, is assigned to alternating stacking of PEDOT and PSS(*36, 37*). Upon hydration, this spacing expands to ~21.72 Å, consistent with previous X-ray diffraction studies(*38, 39*). A weaker feature associated with PEDOT π–π stacking(*25*) is also observed, shifting slightly from ~3.68 Å in the dry film to ~3.77 Å after hydration. Together, these diffraction results indicate that hydration expands the PEDOT-PSS stacking distance while largely preserving the π-stacking of PEDOT chains. Together with the cryogenic vibrational EELS maps, these diffraction results support a structural picture in which water accumulates preferentially within PSS-rich hydrated regions and expands the polymer microstructure, forming electrolyte-accessible pathways for ion redistribution during homogeneous charge relaxation.

**Molecular engineering of water pathways for bioelectronics**

Building on our finding that water pathways dominate charge relaxation, we sought to physically disrupt these networks through molecular engineering. We designed a polythiophene backbone with bulky, hydrophobic oxytert-butyldimethylsilyl ether (OTBS) sidechains, termed P3AOT-OTBS (**Fig. 5A**, see Materials and Methods for synthesis details). As a hydrophilic control, we obtained P3AOT-OH via a one-step chemical deprotection of the P3AOT-OTBS film. Our hypothesis is that the bulky OTBS groups provide significant steric hindrance(*40*). To penetrate the film, hydrated ions must undergo partial desolvation, thereby minimizing both passive and active swelling(*41*).

Using quartz crystal microbalance with dissipation (QCM-D, **Fig. S9**) measurements showed that P3AOT-OTBS has very low passive swelling of ~0.3%. Despite this low swelling, OTBS-protected polythiophenes still undergo volumetric electrochemical doping (*42*). During doping process, EQCM-D further showed that $Cl^-$ ions entering P3AOT-OTBS lose part of their hydration shell, corresponding to the loss of approximately 2–3 water molecules per ion (**Fig. S10** and schematic in **Fig. 5B**). In contrast, $Cl^-$ ions entering P3AOT-OH bring additional water into the film (**Fig. S10**). These results show that the two side chains place the same polymer backbone in two different hydration regimes: a low water content, ion-trapping regime for P3AOT-OTBS and a high water content, fast-ion transport regime for P3AOT-OH.

This difference strongly affected charge retention. After charging to 0.6 V versus Ag/AgCl, P3AOT-OTBS maintained an open-circuit potential of ~0.53 V after 30 min, whereas P3AOT-OH decayed to ~0.38 V over the same period (**Fig. 5C**). The mass response followed the same trend: P3AOT-OTBS showed little mass loss, while P3AOT-OH lost a significant fraction of its electrochemically stored mass. These results link water-mediated ion motion directly to self-discharge and show that suppressing swelling can stabilize stored ionic charge in aqueous OMIECs.

We next tested whether these contrasting hydration regimes translate into distinct OECT behaviors (**Fig. 5D**). The hydrophilic P3AOT-OH devices exhibited high channel current, greater transconductance, and minimal hysteresis, consistent with rapid water-assisted ion transport (**Fig. 5E**, and **Fig. S11C-D**). In contrast, P3AOT-OTBS exhibited a pronounced hysteresis window of ~0.5 V between the forward and backward transfer sweeps (**Fig 5E** and **Fig. S11A-B**), indicating a substantial retention of the electrochemically injected state. Importantly, this large hysteresis cannot be explained by hydrophobicity alone, as conventional hydrophobic polythiophenes such as P3HT show smaller hysteresis windows (*43*). Instead, the bulky OTBS termination likely introduces an additional steric barrier to hydrated-ion transport, consistent with the EQCM-D measurements mentioned before. The kinetic origin of this behavior is further supported by the scan-rate dependence of the transfer characteristics and by transient measurements: at faster sweeps, the saturation current decreases (**Fig. S12**), and also P3AOT-OH switches rapidly (~1 ms) and P3AOT-OTBS responds more slowly with cumulative current changes under pulses (**Fig. S13**). Together, these results support a picture in which the bulky OTBS side chains hinder hydrated-ion penetration and removal, thereby promoting ion retention within the polymer.

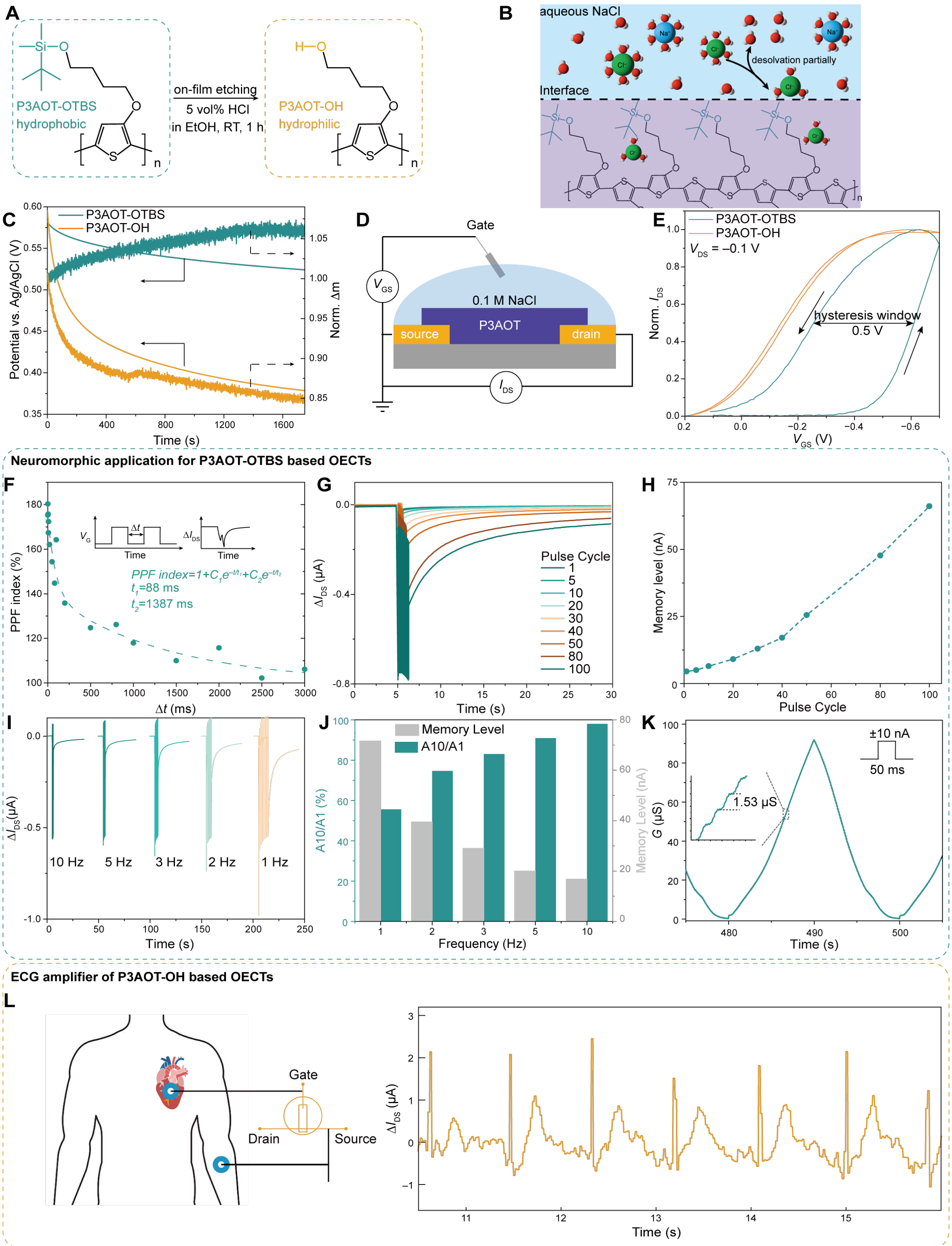

**Fig. 5. Side-chain control of water pathways enables memory and sensing in OMIECs. (A)** Chemical structures of P3AOT-OTBS and its conversion to the hydrophilic P3AOT-OH by one-

step chemical deprotection. **(B)** Schematic showing partial desolvation of hydrated $Cl^-$ ions during penetration into the sterically hindered P3AOT-OTBS film. **(C)** Open-circuit potential retention and mass changes of P3AOT-OTBS and P3AOT-OH films after electrochemical charging. **(D)** Schematic of the P3AOT-based organic electrochemical transistor (OECT). **(E)** Transfer characteristics of P3AOT-OTBS and P3AOT-OH OECTs, showing a large hysteresis window in P3AOT-OTBS. **(F–K)** Neuromorphic response of P3AOT-OTBS OECTs. **(F)** Paired-pulse facilitation as a function of pulse interval Δt. **(G)** Channel-current response to repeated gate-voltage pulses under $V_{DS}$ = –0.1 V and $V_{GS}$ = –0.6 V. (H) Memory level as a function of pulse number. (I) Channel-current response to ten gate pulses applied at different frequencies. **(J)** Memory level and A10/A1 ratio as a function of pulse frequency. **(K)** Long-term potentiation and depression behavior under repeated gate-current pulses, showing gradual and reversible conductance modulation with low programming energy. **(L)** ECG sensing using a P3AOT-OH OECT, showing the device configuration and amplified real-time ECG signal.

Inspired by this ion-retention behavior, we evaluated P3AOT-OTBS OECTs for neuromorphic operation. The devices exhibited history-dependent responses over multiple timescales, consistent with the slow ion transport identified above. Short-term memory was reflected by paired-pulsed facilitation (PPF) of 180% at a 1-ms interval, with fast and slow relaxation components of 88 and 1386 ms (**Fig. 5F**). Repeated gate-voltage pulses stimulation increased the conductance state (**Fig. 5G, H**), while frequency-dependent measurements revealed an inverse trend: increasing the pulse frequency from 1 to 10 Hz reduced the memory level from 71.6 to 16.8 nA, despite an increase in A10/A1 ratio from 57% to 98% (**Fig. 5I, J**). This behavior is consistent with kinetically restricted ion penetration, because hydrated ions have less time to access deeper states at higher frequencies. At longer timescales, repeated gate-current pulses generated 100 continuous and symmetric potentiation/depression states (**Fig. 5K**), with stable cycling and state-dependent programming energies shown in **Fig. S14** and **Fig. S15**. Increasing the programming duration and voltage produced progressively longer-lived conductance states, revealing a transition from short-term to more persistent memory (**Fig. S16**). Together, these results show that restricting hydrated-ion motion converts the same ionic dynamic responsible for relaxation into tunable memory spanning short and long timescales.

To test whether our molecular design can be adapted to different bioelectronics functions, we used the hydrophilic P3AOT-OH polymer-based OECT for real-time electrocardiogram (ECG) recording (**Fig. 5L**). Thanks to swift response and high transconductance, the devices amplified cardiac signals with high temporal fidelity, clearly resolving the main features of the ECG waveform, including the P wave, the QRS complex and the T wave, without noticeable distortion in the temporal $\Delta I_{DS}$ trace.

OMIECs are uniquely suited for bioelectronics because they can seamlessly translate between biological ionic signals and electronic hardware signals. Establishing clear structure-property relationships in OMIECs is essential for rational material design and device optimization. However, this becomes highly complex when these materials operate in physiological environments. In this work, using operando optical microscopy, we observed a homogeneous charge relaxation phenomenon. We attribute this behavior to water pathways, which are introduced by hydrophilic sidechains and the penetration of hydrated ions. We validated this mechanism using the Nernst-Planck model, alongside structural and elemental mapping via cryo-STEM incorporated with EELS and SED. Finally, as a proof of concept, we demonstrated that simply changing the terminal side-chain group can engineer the water content within the film. These molecular tuning redirects the identical conjugated backbone into two distinct operating regimes: steric ion-trapping for robust energy storage and synaptic memory, or unhindered ion transport for high-speed physiological sensing.

Water has long been recognized as a double-edged component in OMIEC devices, but its role has primarily been considered in the context of operational stability. Hydrophilic side chains have become a central design principle for bioelectronics-oriented OMIECs, because hydration accelerates ion transport, enhances ionic-electronic coupling, and enables rapid and sensitive device responses in physiological environments. Our results suggest, however, that hydration should also be considered as a determinant of device function. Hydrated OMIECs naturally favor fast and reversible operation, but the same water-rich pathways facilitate ion escape from the polymer matrix, making such materials poorly suited to bioelectronic functions that require persistent ionic charge storage or memory. This shifts the role of hydration from a parameter to be optimized for stability and speed to a molecular design variable that selects between volatile and persistent bioelectronic functions. Simply replacing hydrophilic OMIECs with conventional hydrophobic semiconducting polymers does not solve this problem: in materials such as P3HT, ion penetration is strongly restricted, and trapping is largely confined near the film surface. By introducing steric hindrance via a bulky and branched hydrophobic termination group, we offer a promising strategy to overcome this limitation. This specific molecular design forces the partial desolvation of hydrated ions, enabling deep and stable bulk trapping. Future work should focus on exploring the delicate balance between rational chemical design and targeted device engineering.

**Acknowledgments:** We thank Dr. Yifei Luo, Dr. Dongxun Lyu and Prof. Clare Grey for fruitful discussions. **Funding:** We acknowledge the funding from the Engineering and Physical Sciences Research Council (EP/W017091/1). S. Z. acknowledges funding from the Swedish Research Council, Vetenskapsrådet 2023-06507 hosted by Linköping University. J.T. and I.M. acknowledge financial support from EU Horizon 2020 grant agreement n°952911, BOOSTER, as well as EPSRC Projects EP/Z536258/1, UKRI2634, and EP/X038777/1. X.W.C. acknowledges funding from the Agency for Science, Technology and Research (A*STAR, Singapore) under the National Science Scholarship. G. E. L., S. M. F and C. D. acknowledge the Wolfson Electron Microscopy Suite and the use of the Thermo Fisher Spectra 300 TEM funded by EPSRC under grant EP/R008779/1. For the purpose of open access, the author has applied a Creative Commons Attribution (CC BY) licence to any Author Accepted Manuscript version arising.

**Author contributions:** S.Z., A. R., and G. G. M. conceived the work. S.Z. performed the device fabrication, characterization and data analysis, as well as the COMSOL modeling; S.T.K built the CP setup which maintained by S. Z. and Z. Q. under supervision of H. S.; S. Z. and G. E. L. prepared the samples for cryo-EM; G. E. L. and S. M. F. collected the cryo-EM. data under supervision of C. D.; S. M. F. processed the cryo-EM data; J. T. designed and synthesized P3AOT-OTBS polymers; C. K. synthesized p(g3T2-T); I. M. supervised J. T. and C. K.; C. S. helped the ECG measurements; A. R. and G. G. M. supervised the project; S. Z., A. R. and G. G. M. wrote the manuscript with review & editing from all authors.

**Competing interests:** Authors declare that they have no competing interests.

**Data, code, and materials availability:** The data underlying all figures in this article are publicly available from the University of Cambridge repository at (DOI to be added in proofs). Code is available upon request from the corresponding authors.

**Supplementary Materials**

Materials and Methods

Supplementary Text 1

Figs. S1 to S16

References

Movies S1 to S3